\documentclass[12pt]{article}

\usepackage{amsmath,graphicx}
\usepackage{epsf}
\usepackage{graphicx,epsfig}
\usepackage{amsfonts}
\usepackage{amssymb}
\usepackage[nosort]{cite}
\usepackage{setspace}
\usepackage{array}

\def\bk{{\bf k}}

\def\bx{{\bf x}}

\def\CL{{\cal L}}
\def\CO{{\cal O}}

\def\m-{{\mbox{-}}}

\def\half{\frac{1}{2}}

\makeatletter
\renewcommand\section{\@startsection {section}{1}{\z@}%
                                 {-3.5ex \@plus -1ex \@minus -.2ex}%
                                   {2.3ex \@plus.2ex}%
                                   {\normalfont\large\bfseries}}
\renewcommand\subsection{\@startsection{subsection}{2}{\z@}%
                                   {-3.25ex\@plus -1ex \@minus -.2ex}%
                                     {1.5ex \@plus .2ex}%
                                     {\normalfont\bfseries}}
\renewcommand\subsubsection{\@startsection{subsubsection}{3}{\z@}%
                                   {-3.25ex\@plus -1ex \@minus -.2ex}%
                                     {1.5ex \@plus .2ex}%
                                     {\normalfont\itshape}}
\makeatother

\newcommand{\Letter}{
\setlength{\textwidth}{16.5cm}
   \setlength{\textheight}{22.6cm}
    \hoffset=-0.5in
\voffset=-2.1cm }

\Letter

\begin{document}
\newcommand{\be}{\begin{equation}}
\newcommand{\ee}{\end{equation}}
\newcommand{\bea}{\begin{eqnarray}}
\newcommand{\eea}{\end{eqnarray}}
\newcommand{\barr}{\begin{array}}
\newcommand{\earr}{\end{array}}

\thispagestyle{empty}
\begin{flushright}
\end{flushright}

\vspace*{0.3in}
\begin{spacing}{1.1}

\begin{center}
{\large \bf Folded Resonant Non-Gaussianity in
\\[0.3cm] General Single Field Inflation}

\vspace*{0.5in} {Xingang Chen}
\\[.3in]
{\em Center for Theoretical Cosmology, \\
Department of Applied Mathematics and Theoretical Physics, \\
University of Cambridge, Cambridge CB3 0WA, UK
} \\[0.3in]
\end{center}

\begin{center}
{\bf Abstract}
\end{center}
\noindent
We compute a novel type of large non-Gaussianity due to small periodic features in general single field inflationary models. We show that the non-Bunch-Davies vacuum component generated by features, although has a very small amplitude, can have significant impact on the non-Gaussianity. Three mechanisms are turned on simultaneously in such models, namely the resonant effect, non-Bunch-Davies vacuum and higher derivative kinetic terms, resulting in a bispectrum with distinctive shapes and running. The size can be equal to or larger than that previously found in each single mechanism.
Our full results, including the resonant and folded resonant non-Gaussianities, give the leading order bispectra due to general periodic features in general single field inflation.

\vfill

\newpage
\setcounter{page}{1}


\newpage

\section{Introduction}
\setcounter{equation}{0}

Recent progress in the theory of inflationary density perturbations has revealed a variety of mechanisms that can be responsible for large non-Gaussianities \cite{Chen:2010xk}. They are discovered not only in novel models, but also in simple models where such mechanisms were not previously anticipated. For example, while it is certainly expected that small periodic features in an inflation model can give rise to small oscillatory corrections to the power spectrum, it is recently found that the same features can boost the amplitude of the non-Gaussianity by several orders of magnitude \cite{Chen:2008wn,Bean:2008na,Flauger:2009ab,Flauger:2010ja,Hannestad:2009yx}. These features introduce small oscillatory components to the inflationary background. The inflaton quantum modes can resonate with this background while they are still within the horizon, generating large interactions. These models are called the resonance models, and the resulting non-Gaussianity has a very distinct form. In this paper we give another example in which novel large non-Gaussianity can be generated due to periodic features.

Features generically perturb the inflaton away from its Bunch-Davies vacuum state. The mode function of the inflaton is generally a linear superposition of a positive and a negative energy component. The usual Bunch-Davies vacuum is chosen to be the positive one because it asymptotes to the vacuum state of the Minkowski spacetime in the small wave-length limit. However any disturbance due to features, sharp or periodic, generically introduce the other component \cite{Bean:2008na,Flauger:2009ab}. This component has to be small to be consistent with the observed approximately scale-invariant power spectrum. Therefore it can be neglected in the previous computation of the resonant non-Gaussianity.

On the other hand, the negative energy component can enhance the folded bispectra \cite{Chen:2006nt}, or more generally the higher correlation functions containing folded sub-triangles in the momentum configuration \cite{Chen:2009bc}.\footnote{The folded bispectrum means the bispectrum evaluated at the folded triangle configurations ($k_1+k_2-k_3=0$ or cyclic). In higher correlation functions, the folded sub-triangle configurations are defined to be the configurations in which three momenta $k_i$, $k_j$ and $k_{ij}\equiv |\bk_i+\bk_j|$ form a folded triangle, for any $i$, $j$.} With such a component, the phase of the mode function product becomes stationary for the folded momentum triangle, producing a large non-Gaussianity. In Ref.~\cite{Chen:2006nt,Chen:2009bc,Holman:2007na,Xue:2008mk,Meerburg:2009ys}, the simple forms of non-Bunch-Davies vacua are put in by hand, largely motivated by the discussions of the hypothetical trans-Planckian physics \cite{Martin:2000xs}. For example, the inflaton wave-function at the transition point from Bunch-Davies to non-Bunch-Davies vacuum does not solve the equation of motion. Feature models provide solid examples of such vacua with self-consistent forms. In the resonance models we consider in this paper, although the amplitude of the non-Bunch-Davies vacuum is very small, it is generated deep within the horizon. As we will see, this dramatically enhances the folded non-Gaussianity through the interaction terms in the action that are least suppressed by slow-variation parameters. These interaction terms can be different from those important for the resonant non-Gaussianity.

We comment that the type of non-Gaussianity we compute here arises not because we introduce new ingredients in the model. It exists in known types of models but previously unnoticed. This was pointed out in Ref.~\cite{Chen:2010xk}. The effect will turn out to be most dramatic in the framework of general single field inflation models \cite{Chen:2006nt}. So we consider the inflaton Lagrangian as a general function of the inflaton $\phi$ and its first derivative \cite{Garriga:1999vw},
\bea
\CL= P(X,\phi) ~,
\eea
where $X\equiv -\half g^{\mu\nu} \partial_\mu \phi \partial_\nu \phi$. Small periodic features can come from different places, for example, the potential for slow-roll inflation \cite{Bean:2008na,Flauger:2009ab} or the warp factor in the kinetic term for DBI inflation \cite{Bean:2008na}. To be general, irrespective of the mechanisms, the only quantities that we will keep track of are its oscillatory frequency $\omega$ and some general oscillation amplitudes that we will parameterize later.
Other notations are the same as in \cite{Chen:2006nt}. The background metric is
\bea
ds^2= -dt^2 + a^2 d\bx^2 ~.
\eea
The three slow-variation parameters are
\bea
\epsilon = -\frac{\dot H}{H^2} ~, \quad
\eta = \frac{\dot\epsilon}{\epsilon H} ~, \quad
s = \frac{\dot c_s}{c_s H} ~.
\eea
The two parameters that describe the higher derivative kinetic terms are
\bea
c_s^2 = \frac{P_{,X}}{P_{,X}+2XP_{,XX}} ~, \quad
\frac{\lambda}{\Sigma} = \frac{ 3XP_{,XX} + 2 X^2 P_{,XXX}}{3P_{,X} + 6X^2 P_{,XX}} ~,
\eea
where subscript "$X$" denotes the derivative with respect to $X$.
Both $c_s$ and $\lambda/\Sigma$ are kept arbitrary.
The scalar perturbation $\zeta$ comes in when we perturb the scale factor as $a^2 e^{2\zeta}$ in the comoving gauge in which the scalar field has no fluctuation. It is conserved after horizon exit and is related to the CMB temperature fluctuations by $\zeta \approx -5 \Delta T/T$. We will calculate its correlation functions.
We will write the bispectrum in terms of a function $S(k_1,k_2,k_3)$ defined as follows \cite{Chen:2010xk},
\bea
\langle \zeta^3 \rangle \equiv S(k_1,k_2,k_3) \frac{1}{(k_1k_2k_3)^2} \tilde P_\zeta^2 (2\pi)^7 \delta^3 ( \sum_{i=1}^3 \bk_i ) ~,
\label{3pt_def}
\eea
where $\tilde P_\zeta = H^2/(8\pi^2 \epsilon c_s)$ is the leading order power spectrum.

\section{Resonant non-Gaussianity}
\setcounter{equation}{0}

We first compute the resonant non-Gaussianity \cite{Chen:2008wn} in the present context of the general single field models.
For this purpose, the most important terms in the cubic action of the scalar perturbation \cite{Chen:2006nt} are
\bea
S_3 \supset \int dt d^3x \left[
\frac{a^3 \epsilon}{2 c_s^2} \frac{d}{dt}\left( \frac{\eta}{c_s^2} \right) \zeta^2 \dot\zeta
-\frac{2a\epsilon s}{c_s^2} \zeta (\partial \zeta)^2
+ \frac{a^3\epsilon}{Hc_s^2} \left( \frac{1}{c_s^2}-1
-\frac{2\lambda}{\Sigma}\right) \dot\zeta^3
\right]
~.
\label{S3_feat}
\eea
This is because these terms contain either the highest time derivatives of the background parameters $H$ and $c_s$, or the least number of factors of $a$ (in terms of conformal time $d\tau= dt/a$ and $\zeta'=a \dot\zeta$), both contribute factors of $\omega/H$. The first and third term are more important if the oscillation originates from $\epsilon$, while the second and third term are more important if it is from $c_s$ or $\lambda/\Sigma$.
This gives the following three-point function,
\bea
\langle \zeta^3 \rangle &=& i \left( \prod_i u_{k_i}(0) \right)
\int_{-\infty}^0 d\tau
\left[
\frac{a^3\epsilon}{c_s^2} \frac{d}{dt} \left( \frac{\eta}{c_s^2} \right) ~
u_{k_1}^* u_{k_2}^* \frac{d u_{k_3}^*}{d\tau}
+ \frac{4a^2 s \epsilon}{c_s^2} (\bk_2\cdot \bk_3) \prod_i u_{k_i}^*(\tau)
\right.
\nonumber \\
&+& \left. \frac{2 a \epsilon}{H c_s^2} \left( \frac{1}{c_s^2}-1-\frac{2\lambda}{\Sigma} \right) \prod_i \frac{d u_{k_i}^*}{d\tau}
\right]
~ (2\pi)^3 \delta(\sum_i \bk_i) + {\rm 2~perm.} + {\rm c.c.} ~,
\label{3pt_res}
\eea
where the ``2~perm." denotes two other terms with permutation of indices 1, 2 and 3. The first term is a generalization of that used in \cite{Chen:2008wn}. But in general single field inflation, the other two terms can also be the leading terms.
The dominant component in the mode function $u_k$ is the positive energy mode
\bea
u_k(\tau) = \frac{i H}{\sqrt{4\epsilon c_s k^3}} (1+ i k c_s \tau) e^{-i k c_s \tau} ~.
\label{mode_BD}
\eea
We separate the coupling into a non-oscillatory part and an oscillatory part, and denote the amplitude of the latter with a subscript $A$. For example,
\bea
\frac{d}{dt}\left( \frac{\eta}{c_s^2} \right) \equiv
\frac{d}{dt} \left( \frac{\eta}{c_s^2} \right)_0 +
\frac{d}{dt} \left( \frac{\eta}{c_s^2} \right)_A \sin(\omega t) ~,
\label{coupling_amp}
\eea
and similarly for the others. We assume that all the oscillatory parameters have the same frequency and phase, otherwise simple modifications are necessary, especially for the phase.

The case with non-oscillatory couplings and Bunch-Davies vacuum (\ref{mode_BD}) in general single field models is considered in \cite{Chen:2006nt}. In this section we focus on the oscillatory part, and in the next section we study the effect of the non-Bunch-Davies correction.

In resonance models, a common integral that we encounter when we compute the non-Gaussianity or wave function of mode function is of the form
\bea
\int_{-\infty}^{\tau} d\tau e^{\pm i\omega t} e^{\mp i K c_s \tau} f(\tau) ~.
\label{res_int}
\eea
This integral can be interpreted as follows \cite{Chen:2008wn}.
The first oscillatory factor is from the couplings such as (\ref{coupling_amp}) and has the background frequency. The second oscillatory factor is from the mode functions when they are inside the horizon where $K$ is a comoving momentum.
Due to the background periodicity, the phase of the integral repeats after $2\pi H/\omega$ e-fold ($\Delta N_e = - H \Delta t$, $t\to t+2\pi/\omega$, $\tau\to \tau e^{-2\pi H/\omega}$ and $K\to K e^{2\pi H/\omega}$). So the integral must be proportional to $\exp(\pm i(\omega/H) \ln K)$, where the plus/minus sign comes from the fact that a larger $K$ resonates later with a larger $t$, corresponding to a larger/smaller phase as we can see from (\ref{res_int}).
The physical frequency of a $K$-mode decreases as it is stretched by inflation, and the resonance happens when it coincides with $\omega$. So the resonance point $\tau_*$ is given by $d(K c_s \tau)/dt = \omega$, i.e.~$\tau_*= -\omega/(K c_s H)$. The leading contribution of the integration is obtained when each $K$-mode sweeps through this resonance point. We approximate it as a step-like function at $\tau_*$ with a width and an amplitude that we estimate in the following way.
Once the mode frequency differs from $\omega$ by $\Delta\omega$, the integral cancels out if performed over $\Delta t_1 \approx 2\pi/\omega$. On the other hand, it takes $\Delta t_2 \approx \Delta \omega/(\omega H)$ to stretch a mode until its frequency changes by $\Delta\omega$. Equating $\Delta t_1$ and $\Delta t_2$ gives the duration $\Delta\tau \approx -H\tau_* \Delta t \sim \sqrt{\omega/H}/(Kc_s)$, namely the resonance width. It also gives
the number of resonance periods, $\omega \Delta t/(2\pi) \approx \sqrt{\omega/(2\pi H)}$. Since one period of resonance gives $2\pi/(Kc_s)$, the final resonance amplitude is $\sqrt{2\pi \omega/H}/(K c_s)$.
The rest of the integrand, denoted as $f(\tau)$, varies much slower within the resonance width and its contribution can be simply approximated as $f(\tau_*)$.
To summarize, the integral (\ref{res_int}) can be approximated as, up to a $K$-independent phase $\varphi_0$,
\bea
e^{i\varphi_0} \frac{f(\tau_*)}{K c_s} \sqrt{\frac{2\pi\omega}{H}}
~e^{\pm i\frac{\omega}{H} \ln K}
\cdot \half \left[ {\rm Erf} \left( Kc_s \sqrt{\frac{H}{2\omega}} (\tau+ \frac{\omega}{K c_s H}) \right) +1 \right] ~,
\label{int_approx}
\eea
where ${\rm Erf}$ is the error function and used to model the step-like behavior.
The above interpretation becomes the stationary phase approximation when we precisely integrate (\ref{res_int}) using $t = -H^{-1}\ln (-H\tau)$ \cite{Flauger:2010ja}. This gives the precise numerical numbers above and the phase $\varphi_0= \pm (\omega/H)(1-\ln (\omega/c_s)) \pm \pi/4$. It also shows a more complicated behavior with superimposed small oscillations. We will use (\ref{int_approx}) as the leading order approximation.

Large resonance happens only if $\omega \gg H$. The terms with the highest order of $k_i c_s \tau$ dominate in (\ref{3pt_res}). Other terms can be larger in the {\em very} squeezed limit $k_i/k_j < H/\omega$, but they are subleading terms because their contribution to $f_{NL}$'s are suppressed by $\CO(H/\omega)$. We neglect these terms here. Using (\ref{int_approx}) in (\ref{3pt_res}), we get the following bispectrum,
\bea
S_{\rm res} &=& f_{NL}^{\rm res1} \sin \left( \frac{\omega}{H} \ln K + \varphi_{\rm res} \right)
+ f_{NL}^{\rm res2} ~\frac{\sum k_i^2}{K^2} \cos \left( \frac{\omega}{H} \ln K + \varphi_{\rm res} \right)
\nonumber \\
&+& f_{NL}^{\rm res3} ~\frac{k_1 k_2 k_3}{K^3} \sin \left( \frac{\omega}{H} \ln K + \varphi_{\rm res} \right)
~,
\label{S_res}
\eea
where $K\equiv k_1+k_2+k_3$, $\varphi_{\rm res}= -\pi/4 +(\omega/H)(1-\ln(\omega/c_s))$ and
\bea
f_{NL}^{\rm res1} &=& \frac{\sqrt{\pi}}{8\sqrt{2}} ~\frac{1}{H} \frac{d}{dt} \left( \frac{\eta}{c_s^2} \right)_A \left( \frac{\omega}{H} \right)^{1/2} ~,
\label{fNL_res1}
\\
f_{NL}^{\rm res2} &=& -\frac{\sqrt{\pi}}{4\sqrt{2}} \frac{s_A}{c_s^2} \left( \frac{\omega}{H} \right)^{3/2} ~,
\label{fNL_res2}
\\
f_{NL}^{\rm res3} &=& -\frac{3\sqrt{\pi}}{4\sqrt{2}}
\left[ \left( \frac{1}{c_s^2} -1 -\frac{2\lambda}{\Sigma} \right) \frac{\epsilon}{c_s^2} \right]_A
\left( \frac{c_s^2}{\epsilon} \right)_0
\left( \frac{\omega}{H} \right)^{5/2} ~.
\label{fNL_res3}
\eea
The first component is a generalization of \cite{Chen:2008wn} and has the same running behavior. The second has an additional mild shape. The third has an additional equilateral shape. All of them can be important depending on the parameters.

\section{Folded resonant non-Gaussianity}

Features also generate a non-Bunch-Davies vacuum component, which creates small oscillatory features in the power spectrum. In the resonance model, long after the mode resonates with the background, the background frequency is too large to have any further significant impact on this mode. Therefore, as in the sharp feature case, we can write the mode function as a linear superposition of two modes \cite{Flauger:2010ja},
\bea
u_k(\tau) = u_+(\tau) + c_- u_-(\tau) ~,
\eea
where $u_+$ is the leading behavior (\ref{mode_BD}) and $u_-$ is the negative energy component,
\bea
u_-(\tau) = \frac{i H}{\sqrt{4\epsilon c_s k^3}} (1-i k c_s \tau) e^{i k c_s \tau} ~.
\eea
The coefficient $c_-$ is small and its time-dependence can be solved using the linear equation of motion,
\bea
v_k'' + c_s^2 k^2 v_k - \frac{z''}{z} v_k = 0 ~,
\eea
where $v_k\equiv z u_k$, $z\equiv a\sqrt{2\epsilon}/c_s$, and
\bea
\frac{z''}{z} = 2 a^2 H^2 \left(1 + \delta \right) ~.
\eea
The $\delta$ denotes
\bea
\delta \equiv -\frac{\epsilon}{2} + \frac{3\eta}{4} - \frac{3s}{2} - \frac{\epsilon\eta}{4} + \frac{\epsilon s}{2} + \frac{\eta^2}{8} - \frac{\eta s}{2} + \frac{s^2}{2} + \frac{\dot \eta}{4H} - \frac{\dot s}{2H} ~,
\label{delta_def}
\eea
in which the last two terms are usually the most important for our purpose.
We denote
\bea
\delta \equiv \delta_0 + \delta_A \sin(\omega t) ~,
\label{delta_amp}
\eea
where the second term is the fast oscillatory component and $\delta_A$ denotes its amplitude.
We assume that the oscillation frequency in (\ref{delta_amp}) is the same as that in (\ref{coupling_amp}), as usually happens. We also ignored possible constant phases in (\ref{delta_amp}). Otherwise it is simple to modify correspondingly.
The integration we get from
\bea
\frac{d c_-}{d\tau} \approx - \frac{2\delta_A H}{\omega \tau}  e^{-2ik c_s \tau} \cos(\omega t)
\eea
is of the same type as in (\ref{res_int}), and the $u_-$ component remains approximately zero until it gets excited at the resonance point \cite{Flauger:2009ab,Flauger:2010ja}. Using the approximation (\ref{int_approx}), we get
\bea
c_- \approx \sqrt{2\pi} \delta_A \left( \frac{H}{\omega} \right)^{3/2}
\exp \left[ i\left( \frac{\omega}{H} \ln k + \varphi_- \right) \right]
\cdot \half \left[ {\rm Erf} \left( k c_s \sqrt{\frac{2H}{\omega}} (\tau+ \frac{\omega}{2k c_s H}) \right) +1 \right]
~.
\label{c-component}
\eea
The amplitude of the negative energy mode should be small. So it is negligible in the computation of the resonant bispectrum. However a larger resonance also makes the mode generated deeper within the horizon, at a scale $2H/\omega$ smaller than the horizon. The deeper into the horizon the negative energy component is created the larger the associated folded bispectra \cite{Chen:2006nt}.

Let us look at the details for the case where $1/c_s^2 -1$ or $\lambda/\Sigma$ is equal to or greater than $\CO(1)$. For the purpose of this section, the three most important terms in the cubic action in Ref.~\cite{Chen:2006nt} are
\bea
S_3 \supset \int dt d^3x \left[
\frac{a^3\epsilon}{Hc_s^2} \left( \frac{1}{c_s^2}-1
-\frac{2\lambda}{\Sigma}\right) \dot\zeta^3
- \frac{3a^3\epsilon}{c_s^2} \left(\frac{1}{c_s^2}-1 \right) \zeta\dot\zeta^2
+ a\epsilon \left(\frac{1}{c_s^2}-1\right) \zeta \left(\partial\zeta\right)^2
\right] ~,
\label{action_cubic_single}
\eea
because they are least suppressed by the slow-variation parameters.

We use the first term as an example. We first consider the effect of the cutoff introduced by the step-like function in (\ref{c-component}).
Using $u_-$ in one of the mode functions and considering the leading non-oscillatory part of the coupling, the oscillation factor in the mode product is cancelled for the folded momentum triangle.
Since the $u_-$ component is generated at a scale $\sim H/\omega$ times smaller than the horizon size and the integral is proportional to $\tau^3$, the enhancement is $\sim (\omega/H)^3$. Multiplying the small amplitude in $c_-$ and considering the pre-factor $(1/c_s^2 -1 -2\lambda/\Sigma)$, we get $f_{NL} \sim (1/c_s^2 -1 -2\lambda/\Sigma) \delta_A (\omega/H)^{3/2}$.

The finite width in the step-like function in (\ref{c-component}) is also important. It makes the shape of the bispectrum sharply concentrated near the folded limit. To see this, we note that, away from the folded limit, the mode function product oscillates with a period $\Delta\tau_{\rm phase} \sim (-k_1+k_2+k_3)^{-1} c_s^{-1}$ or its cyclic. If this period is much larger than the width of the step-like function, $\Delta\tau \sim \sqrt{\omega/(2H)}/(k_1 c_s)$, the $u_-$ component can be approximated as being turned on by an infinitely sharp step-function, and the integration around the sharp edge contributes significantly to the bispectrum. But in the opposite limit, the $u_-$ component is turned on very slowly, and the fast oscillation from the mode functions averages out the effect of the step. Therefore, if we use the quantity $-1+k_2/k_1+k_3/k_1$ (and its cyclic) to describe the deviation from the folded triangle limit, the bispectrum is suppressed if $-1+k_2/k_1+k_3/k_1 > \sqrt{H/\omega}$ (and its cyclic).

The following is the full leading order results for (\ref{action_cubic_single}). The bispectrum $S$ is a summation of two terms, $S_\lambda$ and $S_c$,
\bea
S_\lambda &=& -\frac{3\sqrt{2\pi}}{32}
\left( \frac{1}{c_s^2}-1-\frac{2\lambda}{\Sigma} \right)
\delta_A \left( \frac{\omega}{H} \right)^{3/2}
\nonumber \\
&\times& \frac{k_2 k_3}{k_1^2}
\left[ a(x_1) \cos (\frac{\omega}{H} \ln k_1 + \varphi_-)
+ b(x_1) \sin (\frac{\omega}{H} \ln k_1 + \varphi_-) \right]
+ {\rm 2~perm.} ~,
\label{S_fold_res_lambda}
\\
S_c &=& \frac{\sqrt{2\pi}}{16} \left( \frac{1}{c_s^2}-1 \right)
\delta_A \left( \frac{\omega}{H} \right)^{1/2}
\nonumber \\
&\times& \sum_{i,j,k} {\rm coef}(i,j,k) \frac{k_j k_k}{k_i^2}
\left[ c(x_i) \cos (\frac{\omega}{H} \ln k_i + \varphi_-)
+ d(x_i) \sin (\frac{\omega}{H} \ln k_i + \varphi_-) \right] ~,
\label{S_fold_res_c}
\eea
where ${\rm coef}(i,j,k)=3/2 ~ (j\ne k\ne i)$, $-3/2 ~ (j\ne k, j({\rm or}~ k)=i)$, $1/2 ~ (j=k)$, and we sum over all 27 permutations. We have also defined $x_1\equiv (1-k_2/k_1-k_3/k_1)(\omega/2H)$ and cyclic, and
\bea
a(x) &=& \frac{1}{x^3} \left[ 2+ (x^2-2-\frac{H}{\omega}x^2) ~e^{-\frac{H}{2\omega} x^2} \cos x - (2x + \frac{2H}{\omega} x^3) ~e^{-\frac{H}{2\omega} x^2} \sin x \right] ~,
\label{abcd1}
\\
b(x) &=& \frac{1}{x^3} \left[ (x^2-2-\frac{H}{\omega}x^2) ~e^{-\frac{H}{2\omega} x^2} \sin x + (2x + \frac{2H}{\omega} x^3) ~e^{-\frac{H}{2\omega} x^2} \cos x \right] ~,
\\
c(x) &=& \frac{1}{x^2} \left( 1- e^{-\frac{H}{2\omega} x^2} \cos x - x e^{-\frac{H}{2\omega} x^2} \sin x \right) ~,
\\
d(x) &=& \frac{1}{x^2} \left( - e^{-\frac{H}{2\omega} x^2} \sin x + x e^{-\frac{H}{2\omega} x^2} \cos x \right) ~.
\label{abcd4}
\eea
These functions are the leading terms for $\omega/H\gg 1$. Note that (\ref{abcd1})-(\ref{abcd4}) are all finite at the folded limit $x\to 0$.
The phase $\varphi_-$ is related to $\varphi_{\rm res}$ in (\ref{S_res}) by $\varphi_-=\varphi_{\rm res}+ (\omega/H)\ln 2 + \pi/2$.
For $S_c$, we integrated over the terms with the highest order of $k_i c_s \tau$. The other terms are the subleading contribution which can become important in the very squeezed limit ($k_i/k_j < H/\omega$).
It is easy to see that this bispectrum is periodic-scale-invariant, under $k_i \to k_i \exp(2\pi n H/\omega)$ where $n$ is an integer, due to the symmetry of the Lagrangian.

\begin{figure}[t]
\begin{center}
\begin{tabular}{m{7.5cm} m{7.5cm}}
\epsfig{file=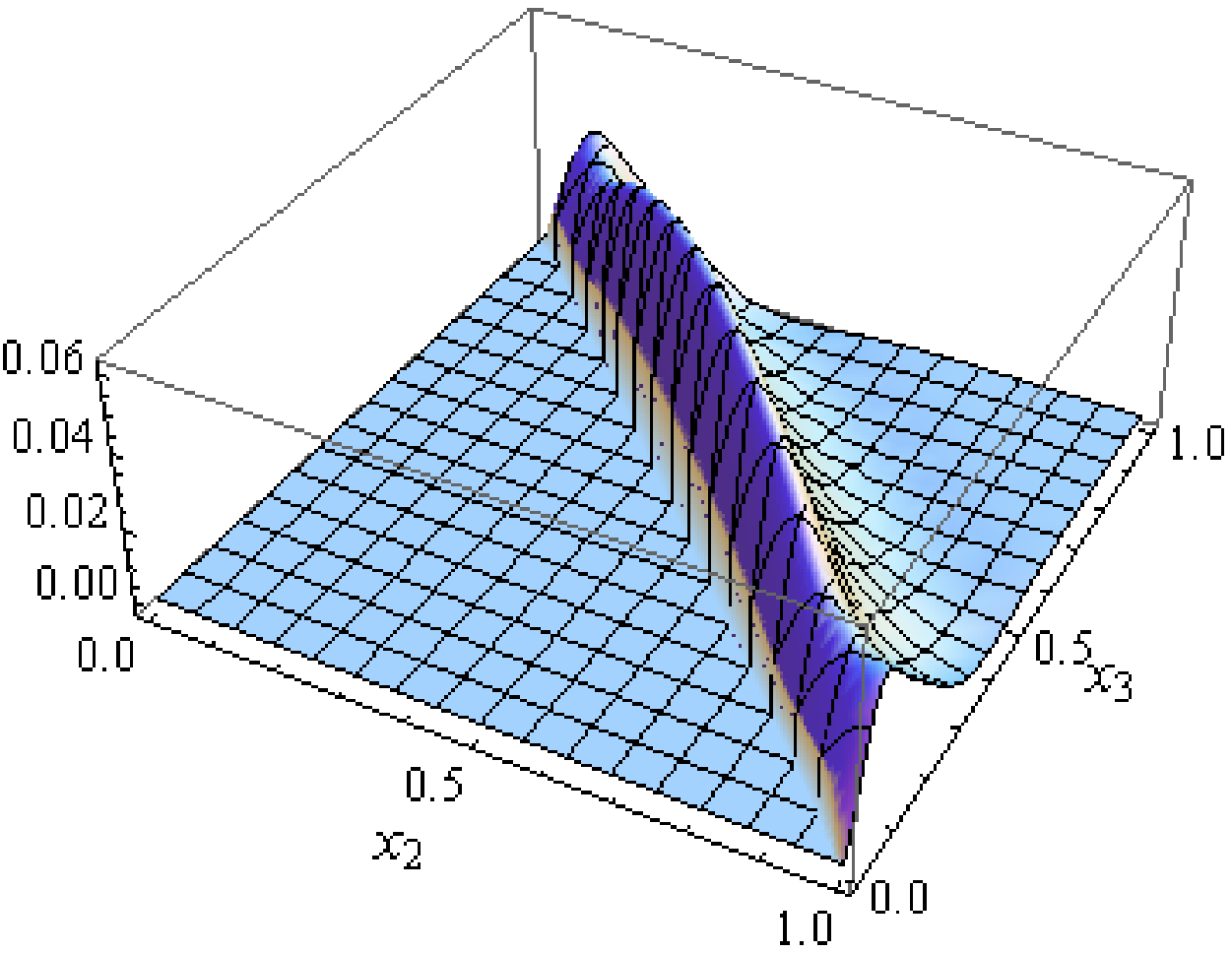, width=0.45\textwidth} &
\epsfig{file=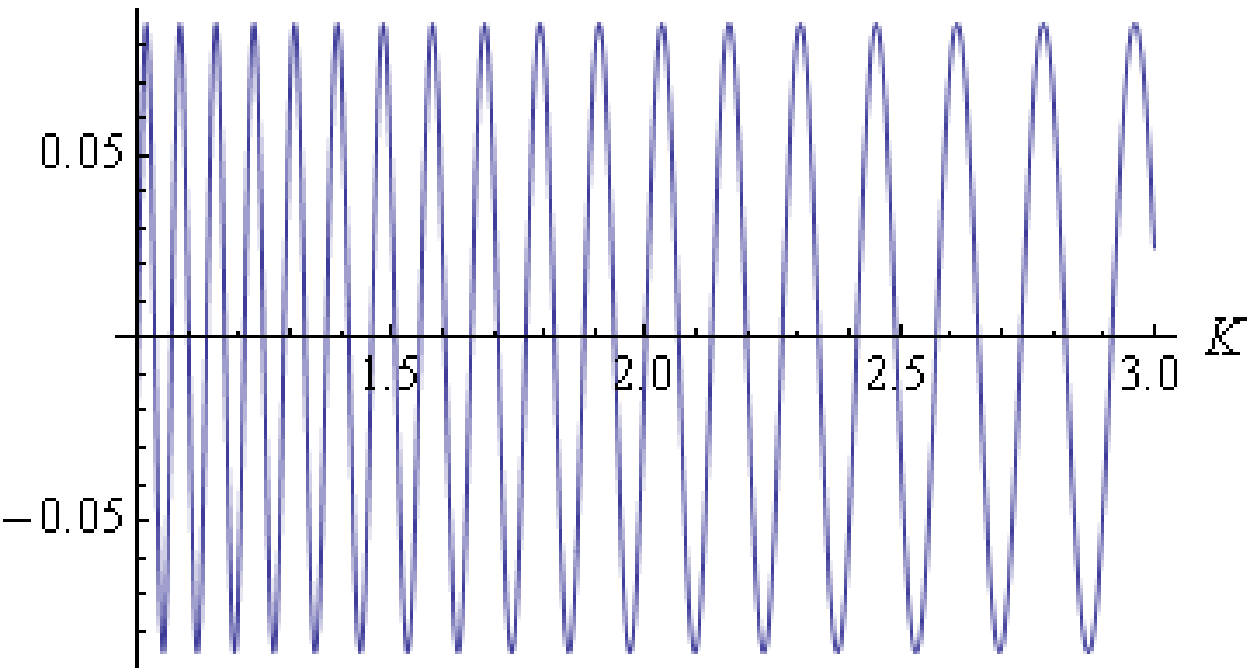, width=0.45\textwidth}
\end{tabular}
\end{center}
\vspace{-0.5cm}
\caption{Shape and running of a folded resonant bispectrum $S_\lambda$: the 2nd line of (\ref{S_fold_res_lambda}) with $\omega/H=100$, $\varphi_-=0$ and $K=3$ (for the shape).}
\label{Fig:fold_res_Sl}
\end{figure}

\begin{figure}[ht]
\begin{center}
\begin{tabular}{m{7.5cm} m{7.5cm}}
\epsfig{file=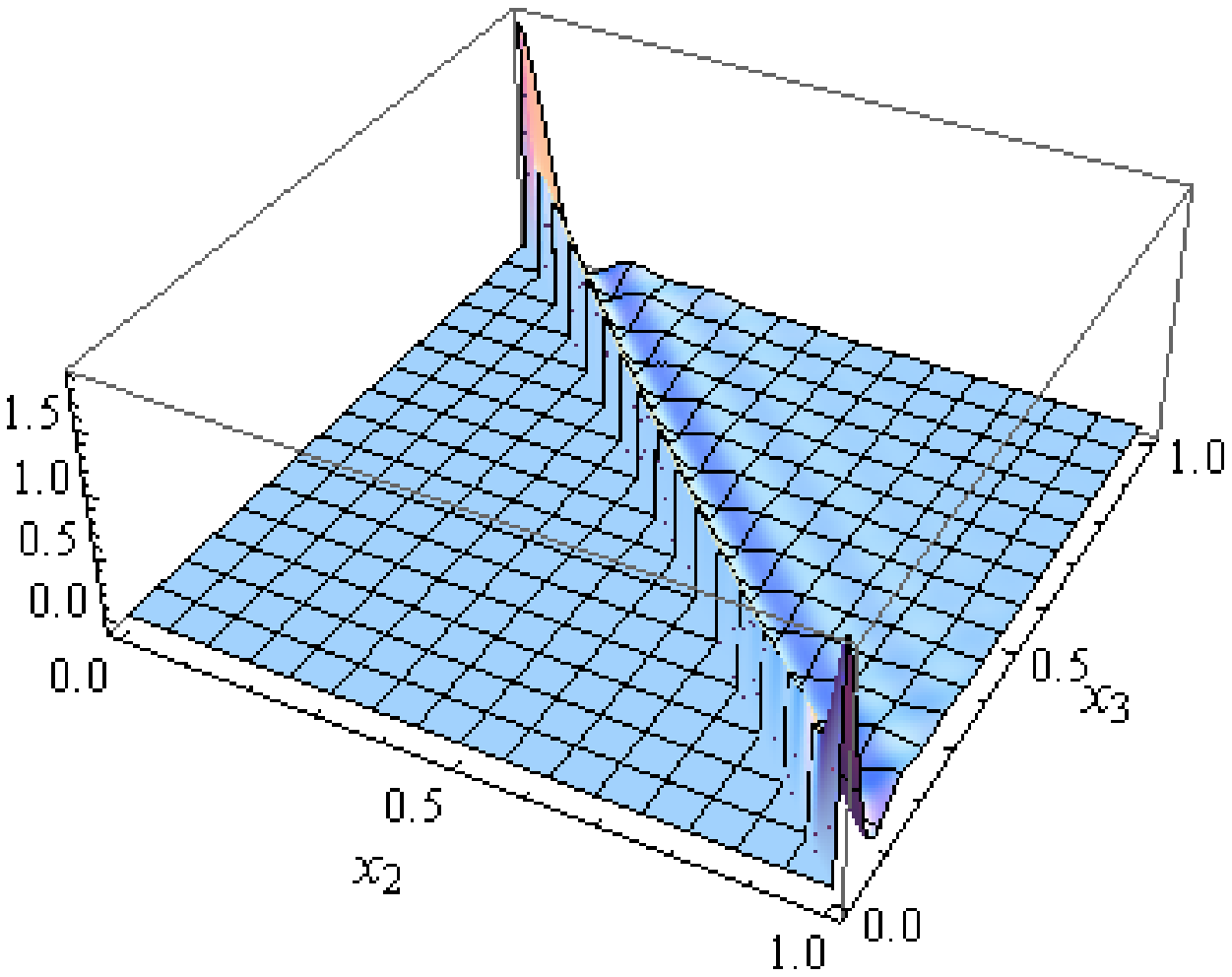, width=0.45\textwidth} &
\epsfig{file=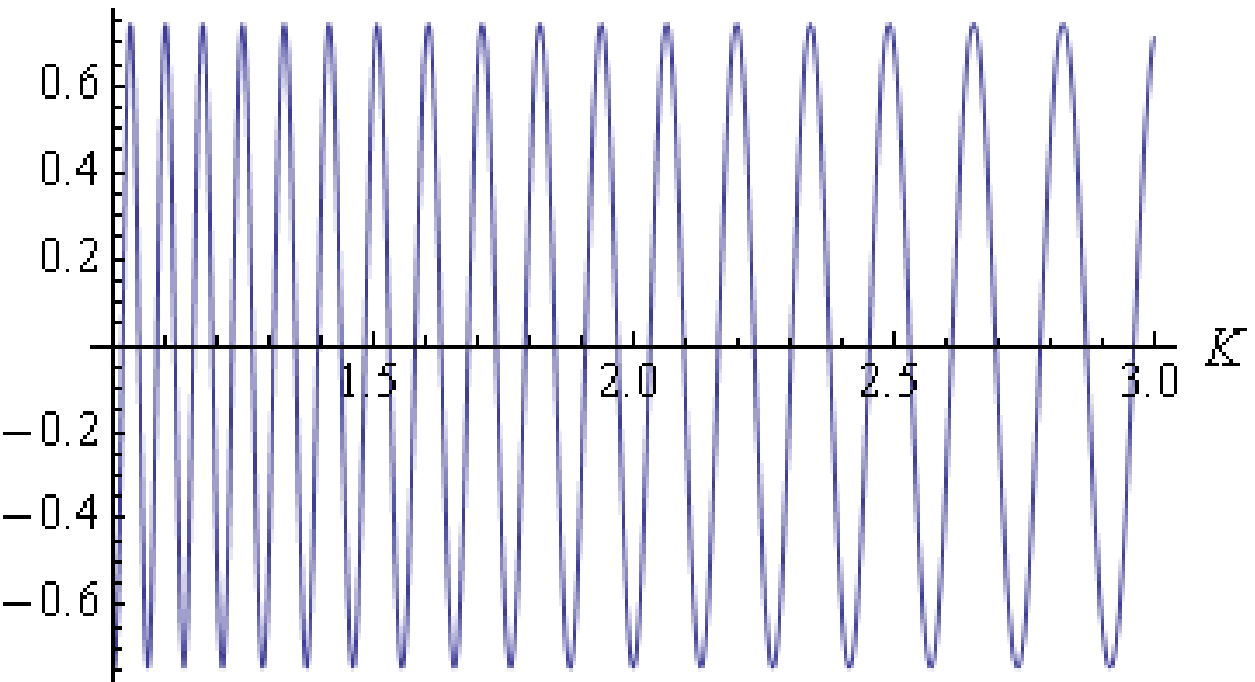, width=0.45\textwidth}
\end{tabular}
\end{center}
\vspace{-0.5cm}
\caption{Shape and running of a folded resonant bispectrum $S_c$: the 2nd line of (\ref{S_fold_res_c}) with $\omega/H=100$, $\varphi_-=\pi$ and $K=3$ (for the shape).}
\label{Fig:fold_res_Sc}
\end{figure}

Since both the shape and running of this bispectrum are important, instead of plotting $S(1,k_2,k_3)$ as usual, we plot the shape and running separately. The shape is the dependence on the momenta ratio $x_2\equiv k_2/k_1$ and $x_3 \equiv k_3/k_1$ with fixed $K=k_1+k_2+k_3$. So we plot $S(K/(1+x_2+x_3),K x_2/(1+x_2+x_3),K x_3/(1+x_2+x_3))$. The running is the dependence on $K$ with a fixed shape which we choose to be $k_1=2k_2=2k_3$. So we plot $S(K/2,K/4,K/4)$. The folded shape and the resonant running that we have discussed qualitatively can be clearly seen in Fig.~\ref{Fig:fold_res_Sl} \& \ref{Fig:fold_res_Sc}, and are nontrivially combined. This will become clearer after we describe it with a simple ansatz.

\section{Simple ansatz}

\begin{figure}[t]
\begin{center}
\begin{tabular}{m{7.5cm} m{7.5cm}}
\epsfig{file=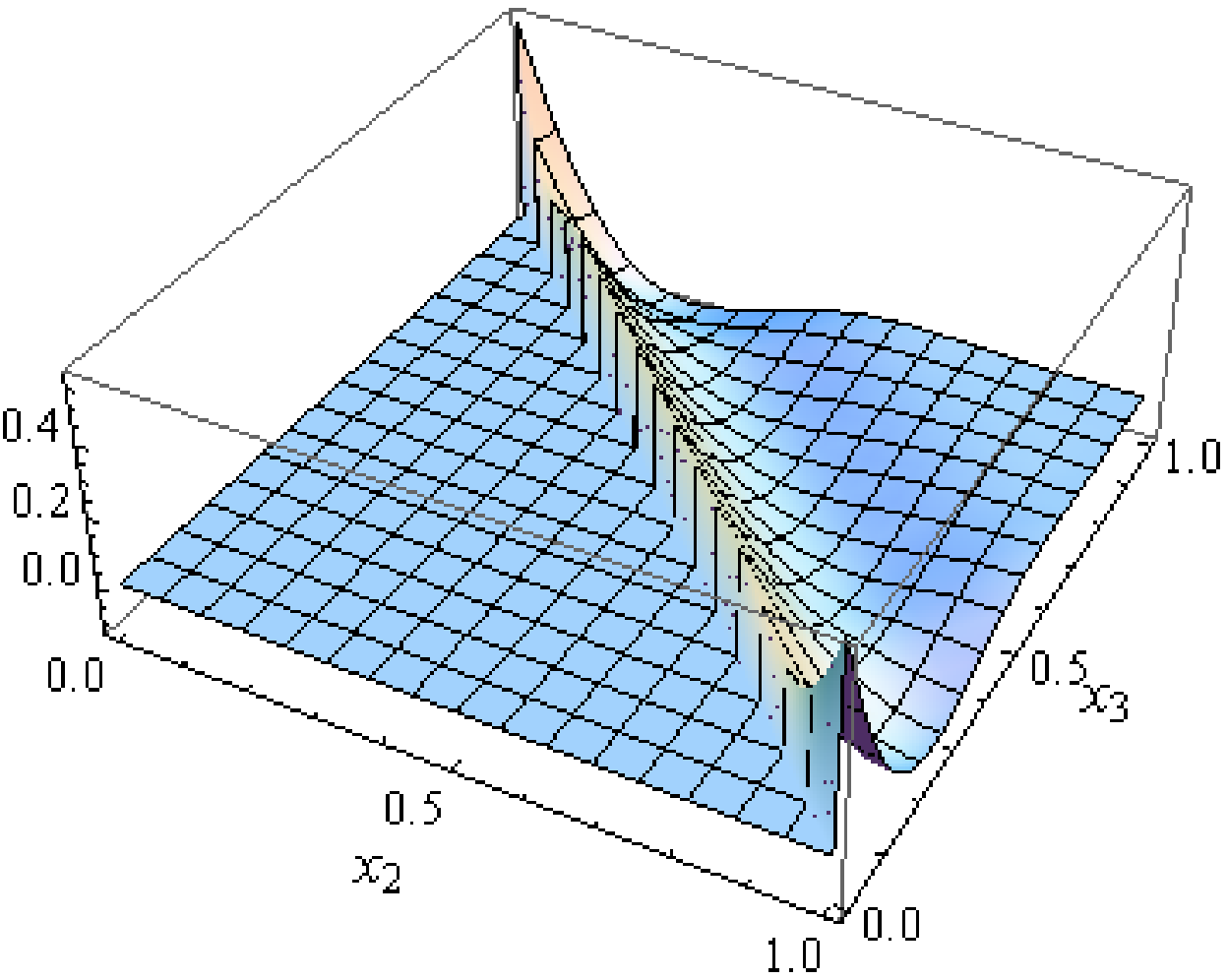, width=0.45\textwidth} &
\epsfig{file=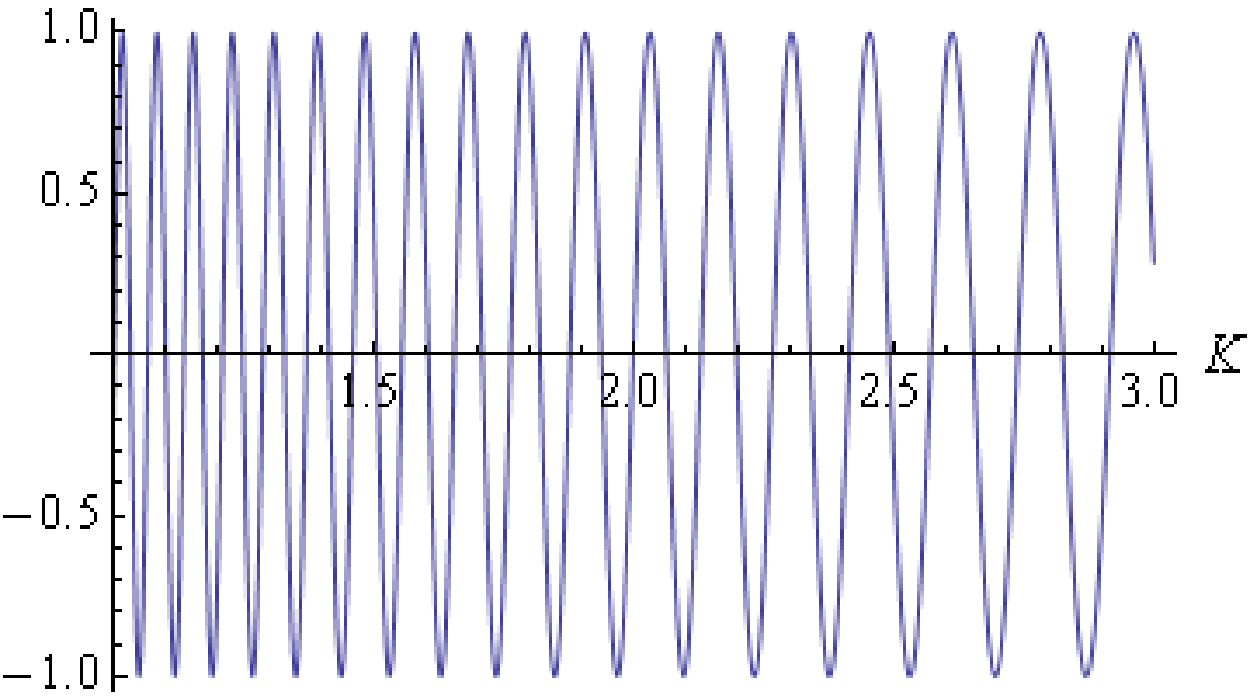, width=0.45\textwidth}
\end{tabular}
\end{center}
\vspace{-0.5cm}
\caption{Shape and running of the ansatz (\ref{ansatz_fold_res}) with $\omega/H=100$, $\varphi=0$ and $K=3$ (for the shape).}
\label{Fig:ansatz_fold_res}
\end{figure}

With the upcoming high precision CMB data, either to test the Gaussianity of the primordial density perturbations or to search for possible large non-Gaussianities, it will be effective to perform general search schemes \cite{Fergusson:2009nv,Regan:2010cn,Fergusson:2010dm,Meerburg:2010ks} followed by targeted optimal analyses \cite{Komatsu:2010fb,Komatsu:2003iq,Creminelli:2006gc,Smith:2009jr}. Simple analytical ansatz of bispectra are necessary for these purposes.
The bispectrum we find here is orthogonal to the scale-invariant folded bispectrum ansatz, and
it is also approximately orthogonal to the resonant bispectrum ansatz (the first term in (\ref{S_res})) due to the leading order non-trivial shape. So a new ansatz is necessary.
We find the most important properties of the bispectrum (\ref{S_fold_res_lambda}) and (\ref{S_fold_res_c}) are captured in the following simple form,
\bea
S_{\rm ansatz}&=& f_{NL}^{\rm fold\m-res}~
\exp\left[-\frac{C^{3/5}}{2} \left( -1 + \frac{k_2}{k_1} + \frac{k_3}{k_1} \right) \right]
\sin \left[ \frac{C}{2} \left( -1 + \frac{k_2}{k_1} + \frac{k_3}{k_1} + 2 \ln k_1 \right) + \varphi \right]
\nonumber \\
&+& {\rm 2~perm.} ~,
\label{ansatz_fold_res}
\eea
where $C=\omega/H$ and the size of $f_{NL}^{\rm fold\m-res}$ is given by the first lines of (\ref{S_fold_res_lambda}) and (\ref{S_fold_res_c}) with proper normalization. An example is shown in Fig.~\ref{Fig:ansatz_fold_res}.
The exponential factor has been chosen to fit the particular decay behavior in (\ref{abcd1})-(\ref{abcd4}) away from the folded limit, namely, $\sim 1/x$ as $1<x<\sqrt{\omega/H}$ and $\sim 1/x^3$ (or $1/x^2$) as $x>\sqrt{\omega/H}$, and may be adjusted.
Such a bispectrum may be constrained or detected using the method of mode decomposition \cite{Fergusson:2009nv,Regan:2010cn,Fergusson:2010dm} by properly choosing the set of base modes.

\section{Size of bispectrum}

From (\ref{c-component}), we know that the correction to the power spectrum is \bea
\delta P_\zeta/P_\zeta = 2\sqrt{2\pi} \delta_A (H/\omega)^{3/2} \cos((\omega/H)\ln k + \varphi_-) ~.
\eea
The periodic oscillation can originate from that in $H$ or $c_s$, and the time derivatives on these parameters contribute factors of $\omega/H$. So to keep $\delta P_\zeta/P_\zeta$ small, we need to keep $(\Delta \epsilon/\epsilon) (\omega/H)^{1/2}$ and $(\Delta c_s/c_s) (\omega/H)^{1/2}$ small, where $\Delta\epsilon$ and $\Delta c_s$ denote the oscillation amplitudes.

On the other hand, the amplitudes of the resonant bispectrum and the folded resonant bispectrum are given by
\bea
f_{NL}^{\rm res} &\sim& \frac{\Delta \epsilon}{\epsilon} \frac{1}{c_s^2} \left( \frac{\omega}{H} \right)^{5/2}
\quad {\rm or} \quad
\frac{\Delta c_s}{c_s} \frac{1}{c_s^2} \left( \frac{\omega}{H} \right)^{5/2}
~,
\\
f_{NL}^{\rm fold\m-res} &\sim& \frac{\Delta \epsilon}{\epsilon} \frac{1}{c_s^2} \left( \frac{\omega}{H} \right)^{7/2 ~{\rm or}~ 5/2}
\quad {\rm or} \quad
\frac{\Delta c_s}{c_s} \frac{1}{c_s^2} \left( \frac{\omega}{H} \right)^{7/2 ~{\rm or}~ 5/2} ~.
\eea
Here $1/c_s^2$ stands for either $1/c_s^2 -1$ or $\lambda/\Sigma$, and $\Delta c_s/c_s$ stands for their relative oscillation amplitudes.
Note that $1/c_s^2-1$ and $\lambda/\Sigma$ do not have to be much larger than one. They can be of order one or somewhat smaller than one, but the $f_{NL}$'s are still large. As we can see, depending on parameters, the folded resonant bispectrum can be the dominant one in the theory, equally important as or more important than the resonant bispectrum (\ref{S_res}) and the equilateral bispectrum in \cite{Chen:2006nt}.
For slow-roll inflation, the terms that we computed for the folded resonant non-Gaussianity are no longer the (only) leading terms. For example, for models with only the canonical kinetic term, a similar calculation with the other terms in the cubic action shows that $f_{NL}^{\rm fold\m-res} \sim \epsilon \delta_A (\omega/H)^{1/2} \sim \Delta\epsilon (\omega/H)^{5/2}$. Comparing to the scale-invariant part $f_{NL} \sim \CO(\epsilon)$, the folded resonant non-Gaussianity can still have an enhanced amplitude, although in this case the resonant non-Gaussianity will be the most dominant component, $f_{NL}^{\rm res} \sim (\Delta\epsilon/\epsilon) (\omega/H)^{5/2}$.

In summary, the overall results of this paper, the resonant non-Gaussianity (\ref{S_res}) and the folded resonant non-Gaussianity (\ref{S_fold_res_lambda}) and (\ref{S_fold_res_c}), give the leading order bispectra due to general periodic features in general single field inflation.

\medskip
\section*{Acknowledgments}

I am supported by the Stephen Hawking advanced fellowship.

\end{spacing}

\end{document}